\documentstyle[aps,12pt,floats,epsfig]{revtex}
\tightenlines 
\begin{document}

\preprint{}

\title{Physics of Antiproton Nuclear Interactions near Threshold}

\author{Avraham Gal} 

\address{Racah Institute of Physics, The Hebrew University,\\ 
Jerusalem 91904, Israel}

\date{\today}
\maketitle 

\begin{abstract}

Antiproton-nucleus optical potentials fitted to $\bar p$-atom 
level shifts and widths are used to calculate the recently 
reported very low energy ($p_{L}<100$ MeV/c) $\bar p$ cross sections 
for annihilation on light nuclei. The apparent suppression of 
annihilation upon increasing the atomic charge $Z$ and mass number 
$A$ is resolved as due to the strong effective repulsion 
produced by the very absorptive optical potential which keeps 
the $\bar p$-nucleus wavefunction substantially outside the 
nuclear surface, so that the resulting reaction cross section 
saturates as function of the strength of Im $V_{{\rm opt}}$. 
This feature, for $E >0$, parallels the recent prediction, 
for $E < 0$, that the level widths of $\bar p$ atoms saturate and,
hence, that $\bar p$ deeply bound atomic states are relatively        
narrow. 
Predictions are made for $\bar p$ annihilation cross sections 
over the entire periodic table at these very low energies and 
the systematics of the calculated cross sections as function of 
$A$, $Z$ and $E$ are discussed and explained in terms of 
a Coulomb-modified strong-absorption model. Finally, optical 
potentials which fit simultaneously low-energy $\bar p - ^4$He 
observables for $E < 0$ as well as for $E > 0$ are used to assess 
the reliability of extracting Coulomb modified $\bar p$ nuclear 
scattering lengths directly from the data. 
\newline
\newline 
Invited talk at the Third International Conference on Perspectives 
in Hadronic Physics, Trieste, May 2001. To appear in Nuclear Physics A. 

\end{abstract}

\section{Introduction}
\label{sec:int}

Antiproton annihilation cross sections at very low energies 
($p_L < 100$ MeV/c) have been reported for light nuclei 
by the OBELIX collaboration 
\cite{Zen99a,Zen99b,Bia00a}. At these energies the total $\bar p$ 
reaction cross section consists only of $\bar p$ annihilation. 
Whereas at relatively higher energies ($p_{L} \approx 200 - 600$ 
MeV/c) the $\bar p$ annihilation cross sections exhibit the 
well known $A^{2/3}$ strong absorption dependence on the nuclear 
target mass number $A$, these cross sections at very low energies 
have defied any simple, obvious regularity. It has been 
demonstrated that the {\it `expected'} $ZA^{1/3}$ dependence on 
the atomic charge $Z$ and mass number $A$ is badly violated 
\cite{Bia00a,Bia00b}. 
Antiproton annihilation cross sections at very low energies 
simply do not rise with $A$ as fast as is anticipated. 

Here I report a recent study of low energy $\bar p$ annihilation 
on nuclei, using the optical model approach \cite{GFB00,BFG01}. 
Optical potentials have 
been very successful in describing strong interaction effects in 
hadronic atoms \cite{BFG97}, including $\bar p$ atoms \cite{BFG95}. 
It has been noted, for pions, that the total reaction cross sections 
at low energies are directly related to the atomic-state widths, and 
that once a suitable optical potential is constructed by reasonably 
fitting it to the atomic level shifts and widths in the negative 
energy bound state domain, these total reaction cross sections are 
reliably calculable \cite{SMC79,FGJ91}. The recent publications 
\cite{Zen99a,Zen99b,Bia00a} of experimental results of total cross 
sections for $\bar p$ annihilation on nuclei at very low energies
raise the intriguing possibility of connecting these two energy
regimes in a systematic way also for antiprotons. However, most
of the data on annihilation cross sections for $\bar p$ 
are for very light nuclei, where the concept of a rather universal
optical potential that depends on $A$ and $Z$ only through the
nuclear densities is questionable. For this reason, in the present 
work, optical potentials are used mostly for crossing the $E=0$ 
borderline within the {\it same} atomic mass range, from bound 
states to scattering. These potentials are strongly absorptive, 
which leads to a remarkable saturation of the total reaction 
cross section with increasing $A$. 

This review is organized as follows. The saturation phenomenon in 
$\bar p$ atoms, and for $\bar p$ total reaction cross sections, 
is described and discussed in Sect. \ref{sec:sat}. Calculational 
results are given, demonstrating the success of using a unified 
optical potential methodology across the $\bar p$-nucleus threshold. 
The $A$ and $Z$ dependence of these low energy 
annihilation cross sections at $p_L=57$ MeV/c is discussed. 
The Coulomb modified strong absorption model at 
{\it very} low energies is described in Sect. \ref{sec:abs} 
and, particularly, how it successfully reproduces, due to 
the Coulomb focussing effect, the optical potential reaction 
cross sections. Lastly, in Sect. \ref{sec:He}, optical potentials 
which fit simultaneously low energy $\bar p - ^4$He observables 
for $E < 0$, as well as for $E > 0$, are used to assess the 
reliability of extracting $\bar p$ nuclear scattering lengths 
directly from the data.

\section{Saturation of $\bar p$ atomic widths and of total reaction 
cross sections}
\label{sec:sat} 

\begin{figure}
\begin{minipage}[t]{75mm}
\epsfig{file=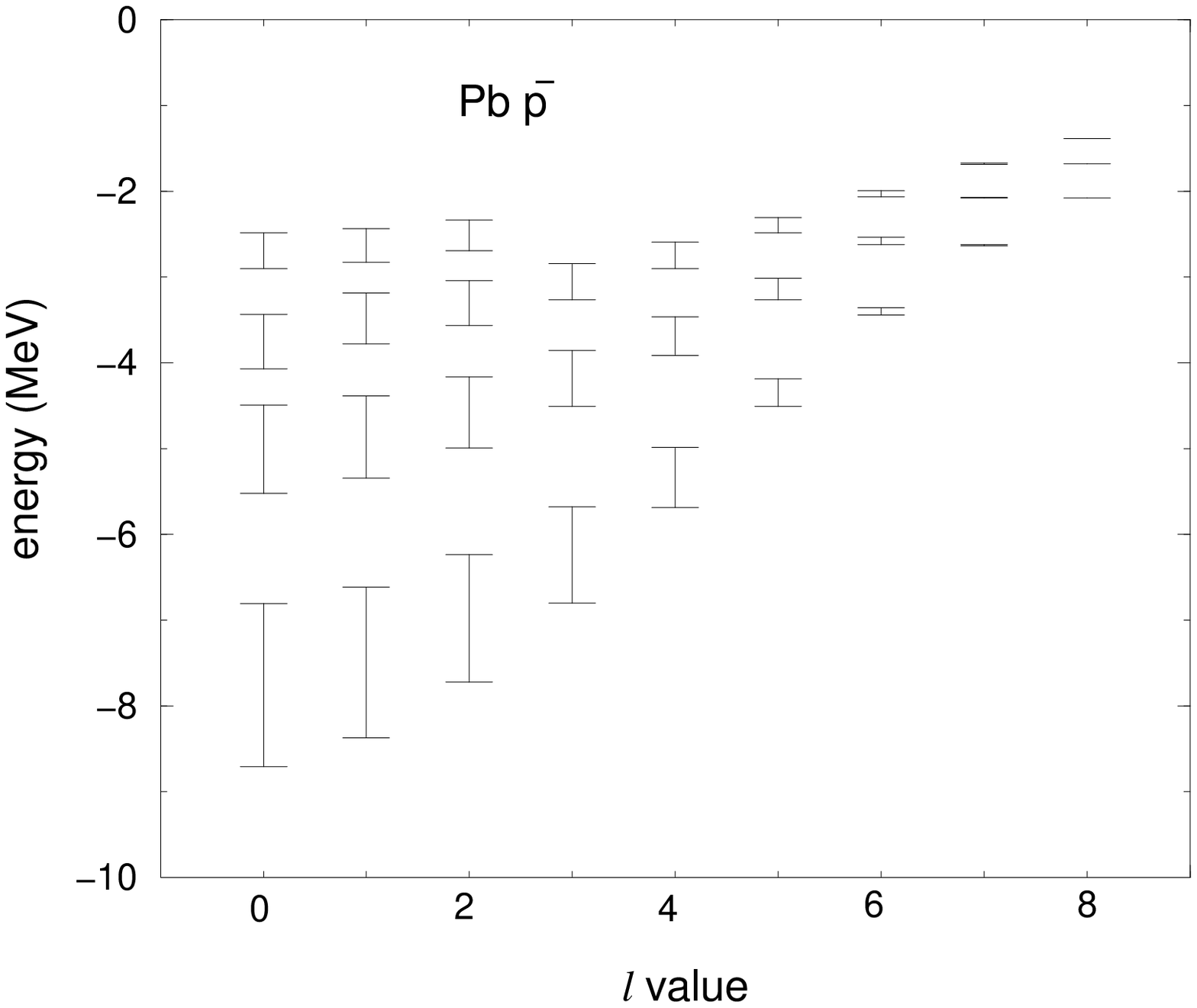,height=90mm,width=75mm} 
\caption{Calculated energy levels for $\bar p$ atoms of Pb.} 
\label{fig:pbarPb} 
\end{minipage} 
\hspace{\fill} 
\begin{minipage}[t]{80mm}
\epsfig{file=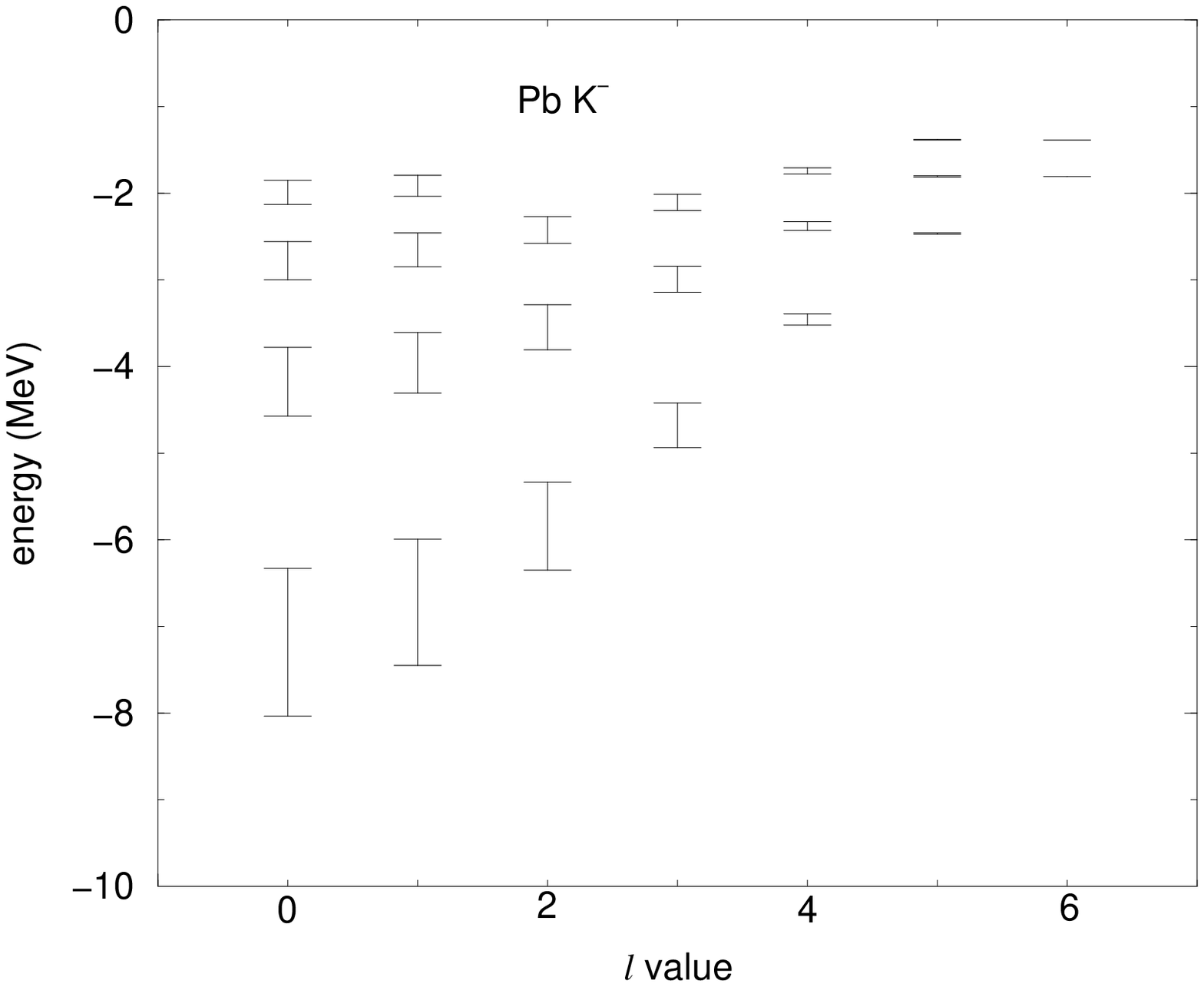,height=90mm,width=80mm} 
\caption{Calculated energy levels for kaonic atoms of Pb.} 
\label{fig:KPb}
\end{minipage}
\end{figure}

Antiprotonic and $K^-$-atom optical potentials are strongly absorptive 
\cite{BFG97}. This strong absorptivity has recently been shown 
\cite{FGa99a,FGa99b} to lead to a saturation 
of the widths of atomic states upon increasing the 
absorptivity of $V_{\rm opt}$, and to the prediction of relatively
narrow deeply bound atomic states. 
Figures \ref{fig:pbarPb} and \ref{fig:KPb} 
show calculated $\bar p$ and $K^-$ atomic energy levels in Pb, 
respectively, for several values of $l$. 
The bars stand for the full width $\Gamma$ of the 
level and the centers of the bars correspond to the 
binding energies Re $B$. It is seen that the calculated widths 
saturate at about 2 MeV. 
The dependence on the model is 
found to be negligibly small, affecting the calculated widths by less than 
$5\%$, provided the optical potential was fitted to the known part of the 
spectrum throughout the periodic table. 
For this heavy nucleus the levels are quite close to each other although
the spectrum is still well defined. The $9k$ $\bar p$ level 
\cite{Trz01} and the $7i$ $K^-$ level \cite{Che75} in Pb 
are the last observed in the respective X-ray spectra. 
The close analogy between bound-state widths and total reaction cross 
sections is best demonstrated, 
assuming for simplicity a Schr\"odinger-type equation, 
by comparing the following expressions 
for the width $\Gamma$ and for the total reaction cross section
at positive energies with each other: 

\begin{equation} \label{equ:sat}
\frac{\Gamma}{2}= -\frac{\int {\rm Im} V_{{\rm opt}}(r) 
| \psi({\bf r}) | ^2  d {\bf r}}
{\int | \psi({\bf r}) | ^2  d {\bf r}}\quad , \quad \quad \quad 
\sigma_R = -\frac{2}{\hbar v} \int | \chi({\bf r}) |^2 
{\rm Im} V_{{\rm opt}}(r) d {\bf r}\quad,
\end{equation}
where $\psi({\bf r})$ is the $\bar p$ full atomic wavefunction 
and $\chi({\bf r})$ is the $\bar p$ - nucleus elastic scattering
wavefunction; $v$ is the c.m. velocity.

A standard $V_{\rm opt}$, to be used in the above expressions,
is given by the `$t\rho$' form \cite{BFG97}

\begin{equation}\label{equ:potl}
2\mu V_{{\rm opt}}(r) =
 -{4\pi}(1+{\frac{A-1}{A}\frac{\mu}{m}})b_0\rho(r) \;\;\; ,
\end{equation}
where $m$ is the nucleon mass, $\mu$ is the reduced $\bar p$-nucleus
mass, $b_0$ is a complex parameter obtained from fits to the data 
and $\rho(r)$ is the nuclear density distribution normalized to $A$.
Although the effect of the $(A-1)/A$ factor may often be absorbed 
into the fitted parameters, as is the case in this section, 
for very light $\bar p$ atoms one must include the effect of the 
underlying $\bar p N$ interaction, at least by folding 
a phenomenological interaction form factor into the nuclear density.
Thus, the $2p$ level shift and width data in $^{3,4}$He \cite{Sch91} 
are well fitted by folding a Gaussian with a range
parameter of 1.4 fm into the nuclear density distribution. 
The resulting potential, with $b_0$=0.49+i3.0 fm, is highly absorptive. 
Using this potential ($a$) to calculate the total reaction
(annihilation) cross section for 57 MeV/c $\bar p$ on $^4$He, 
the calculated cross section is 901 mb, in excellent agreement with the 
reported value of 915$\pm$39 mb \cite{Bia00a}.

\begin{figure} 
\begin{minipage}[t]{75mm}
\epsfig{file=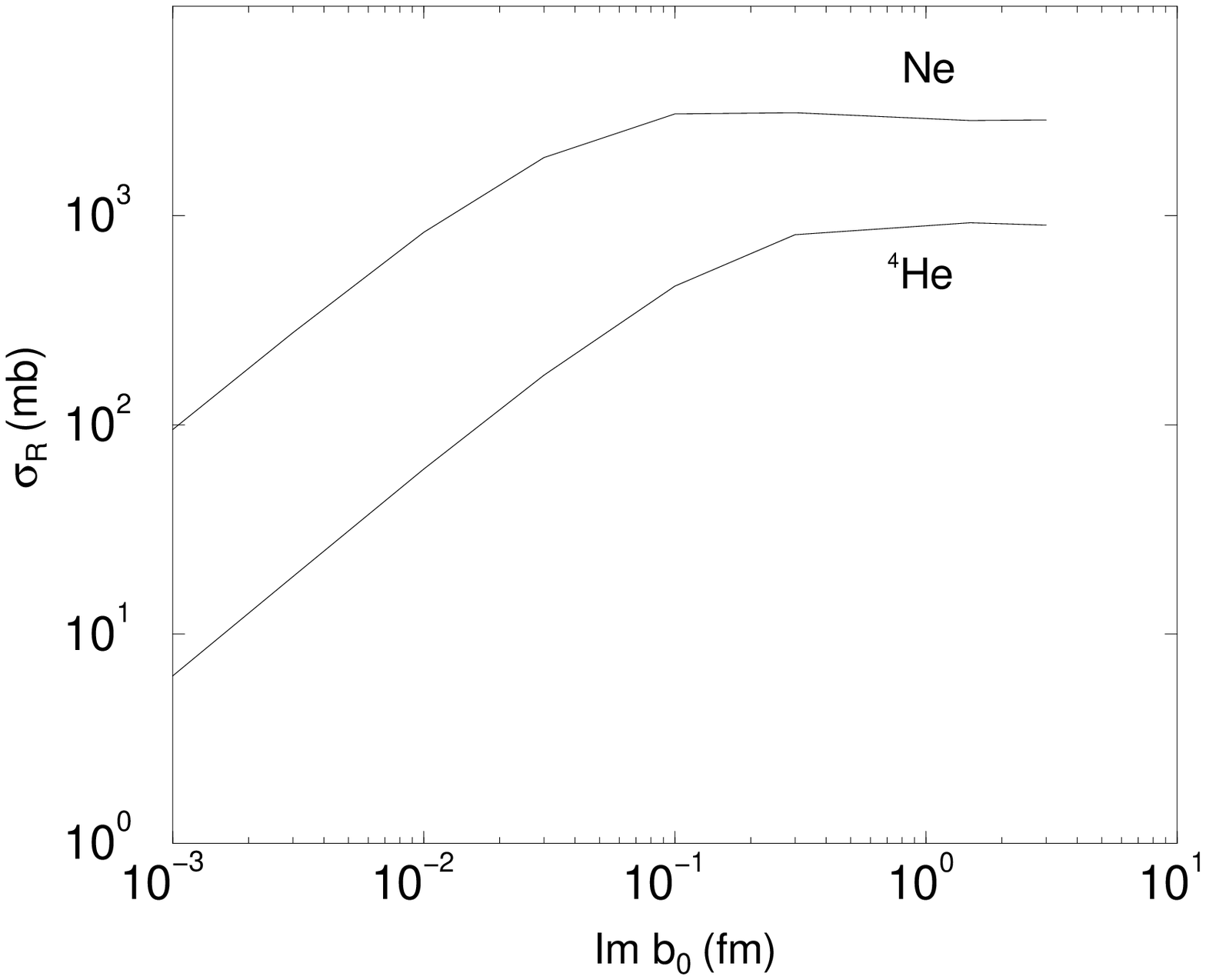,height=85mm,width=75mm} 
\caption{Calculated total reaction cross sections for 
57 MeV/c $\bar p$ on $^4$He and Ne as function 
of the strength parameter Im $b_0$ of the optical potential ($a$).} 
\label{fig:sat} 
\end{minipage} 
\hspace{\fill}
\begin{minipage}[t]{80mm}
\epsfig{file=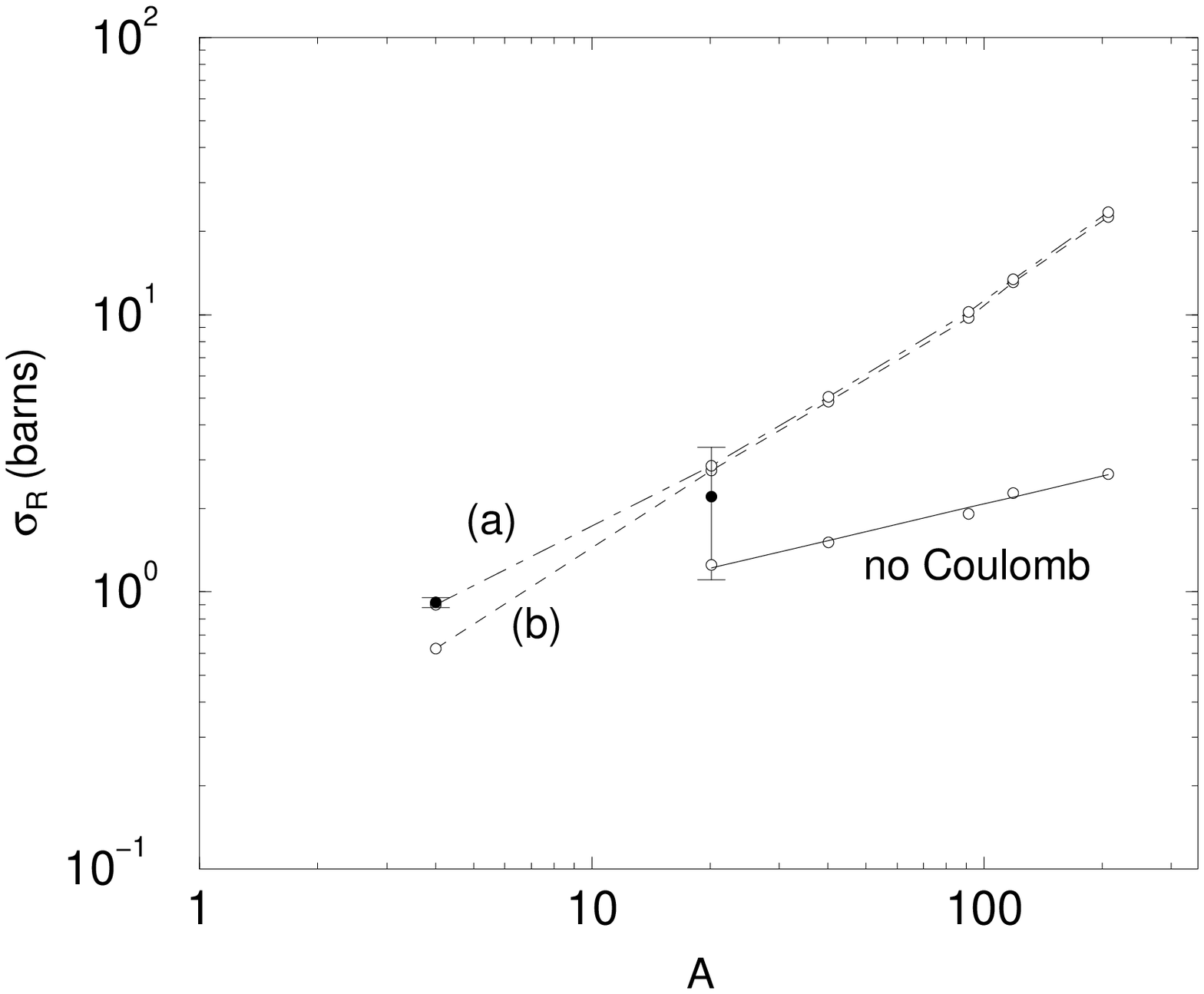,height=85mm,width=80mm} 
\caption{Calculated $\bar p$ total 
reaction cross sections (open circles) at 57 MeV/c 
for potentials ($a$) and ($b$), and for potential ($b$) but 
without the Coulomb interaction.} 
\label{fig:res} 
\end{minipage} 
\end{figure}

Figure \ref{fig:sat} demonstrates the extreme strong absorption
conditions which are relevant to the $\bar p$ nucleus interaction
at very low energies (and for $\bar p$ atoms). It shows calculated
reaction cross sections for $\bar p$ at 57 MeV/c on $^4$He
and Ne as function of the strength Im~$b_0$ of the imaginary part
of the potential ($a$) described above, with the rightmost edge corresponding
to its nominal value Im~$b_0$=3.0 fm. It is seen that as long as the
absorptivity (Im~$b_0$) is very weak, less than 1\% of its nominal value,
$\sigma_R$ is approximately linear in Im~$b_0$, which according to
Eq. (\ref{equ:sat}) means that the $\bar p$ wavefunction depends
weakly on Im~$b_0$. However, already  at below 5\% of the nominal value of
Im~$b_0$ the reaction cross sections begin to saturate, much the same
as for the widths of deeply bound $\bar p$ atomic states \cite{FGa99a,FGa99b}.
The mechanism is the same in both cases, namely exclusion of the wavefunction
from the nucleus due to the absorption, which reduces dramatically the
overlap with the imaginary potential in the integrals of Eq. (\ref{equ:sat}). 
The onset of saturation is determined approximately by the strength parameter
2$\mu$(Im~$V_{\rm opt}$)$R^2$, where $R$ is the radius of the nucleus.
Thus, saturation of $\sigma_R$ in Ne starts at a smaller
absorptivity than its onset in $^4$He. The Ne/$^4$He ratio of $\sigma_R$
values changes, due to this effect, from about 15 in the perturbative
regime of very weak absorptivity to about 3 in the strong absorption regime.
However, the precise, detailed pattern of the change also depends on 
Re~$V_{\rm opt}$ which may enhance or reduce the exclusion of
the $\bar p$ wavefunction from the nucleus.

Figure \ref{fig:res} shows calculated $\bar p$
reaction cross sections at 57 MeV/c across the
periodic table. The dot-dashed line is for the above mentioned
potential ($a$) which is expected to be valid only in the immediate vicinity
of He. The dashed line ($b$) is for the first potential from Table 7 of
Ref. \cite{BFG97} which fits $\bar p$ atom data over the whole periodic
table, starting with carbon. This potential is not expected to
fit data for very light nuclei, and indeed it does not fit the $^4$He
annihilation cross section. However, it is noteworthy that the two potentials 
predict almost the same cross sections for $A > $ 20 and certainly 
display a very similar dependence on $A$. Also shown in the figure is the
recent experimental result \cite{Bia00a} for Ne, with a very limited accuracy.
Furthermore, for $A > $ 20 the points along the solid line are the
calculated $\bar n$ - nucleus total reaction cross sections, obtained
from potential ($b$) by switching off the Coulomb interaction.
The solid line is a fit to an $A^{1/3}$ power law which
appears to be appropriate to strong absorption of uncharged particles 
at very low energies. Comparing the dashed line for 
negatively charged particles with the solid line for uncharged particles, 
it is clear that the $\sigma_R$ values obtained by 
including the Coulomb interaction are considerably enhanced 
at this very low energy with respect to the ones obtained without it. 
This is due to the Coulomb focussing effects which are discussed below. 

Comprehensive measurements of $\bar n$-nucleus total annihilation 
cross sections on six targets from C to Pb at $p_L = 50 - 400$ MeV/c
have been recently reported \cite{Bot01} in terms of a factorized 
dependence on $p_L$ and $A$: 

\begin{equation} \label{equ:nbar} 
{\sigma}_{ann}(p_{L},A)={\sigma}_{0}(p_{L})A^{x} \;,\;\;\; 
x=0.6527 \pm 0.0044 \;, 
\end{equation}
where ${\sigma}_{0}(p_{L})$ is given explicitly as a monotonically 
decreasing function of $p_L$. This essentially uniform $A^{2/3}$ 
behavior, which normally is expected only at considerably higher 
energies, is puzzling. 
Furthermore, the reported $\bar n$ cross sections for a given target, 
particularly at the lower end of the energy range, are too high 
to be compatible with the $\bar p$-atom and $\bar p$-nucleus 
phenomenology considered in this review. For example, 
Eq. (\ref{equ:nbar}) gives about $4000 \pm 600$ mb for $\bar n$ 
annihilation at $p_L$ = 57 MeV/c on Ne, roughly three times the 
value shown in Fig. \ref{fig:res} for $\bar n$-Ne using potential 
($b$), and substantially higher than the $\bar p$-Ne annihilation 
cross section \cite{Bia00a} of $2210 \pm 1105$ mb. This persists, 
although to a lesser extent, also at higher energies. Thus, at 
$p_L$ = 192.8 MeV/c Eq. (\ref{equ:nbar}) gives about $1175 \pm 125$ 
mb for $\bar n$-Ne, whereas the measured annihilation cross section 
for $\bar p$-Ne at this energy is only $956 \pm 47$ mb \cite{Bal86}.

\section{Coulomb focussing effects on total reaction cross sections at 
low energies} 
\label{sec:abs}

The behavior of the calculated $\bar p$-nucleus total reaction 
cross sections at very low energies may be explained in terms of 
a simple Coulomb-modified strong absorption model 
\cite{BFG01} as follows. 
Recall that an attractive Coulomb potential causes focussing of
partial-wave trajectories onto the nucleus, an effect which at low
energies may be evaluated semiclassically, 
as recognized a long time ago by Blair in 
connection with $\alpha$ particle reactions on nuclei \cite{Bla54}
(see Ref. \cite{Schiff} for the applicability of the semiclassical 
approximation to Coulomb scattering). 
Assuming that complete absorption occurs in all partial waves for
which the distance of closest approach is smaller than $R$, one
gets the following relation between the Coulomb-modified $l_{\rm max}$
and $R$:

\begin{equation} \label{eq:lmax}
(l_{\rm max} + \frac {1} {2})^2 \approx (k R)^2
(1 + \frac {2 \eta} {k R}) \;\;\; ,
\end{equation}
where $\eta$ is the standard Coulomb parameter \cite{Schiff}. 
At very low energies, $2\eta >> kR$, and therefore $l_{\rm max} >> kR$
due to the focussing effect of $V_c$. The total reaction cross section
in the strong absorption limit is then given by

\begin{equation}  \label{eq:sigma}
\sigma_R = \frac{\pi}{k^2} \sum (2l+1) \approx \frac{\pi}{k^2}
(l_{\rm max} +\frac{1}{2})^2 \approx \pi R^2 (1 + \frac{2 \eta}{k R})
= \pi R^2 (1 + \frac{2 m Z e^2}{\hbar^2 k_L k R}) \;\;\; ,
\end{equation}
where $k_L$ and $k$ are the laboratory and c.m. wave numbers, respectively.
The second term within the brackets represents the Coulomb
focussing effect and 
at very low energies it becomes dominant \cite{GFB00},
thus leading to a $ZA^{1/3}$ dependence of the cross section
if $R=r_0A^{1/3}$. At high energies, where the Coulomb focussing effects 
become insignificant, the well known strong absorption value 
$\pi R^2$ is obtained, giving rise to the expected $A^{2/3}$ dependence. 
In order to use expression (\ref{eq:sigma}), one needs to define 
an equivalent radius $R$ that will properly represent the 
$\bar p$-nucleus interaction at the particular energy. 
For a density-folded $\bar p$ optical potential which fits the 
atomic data for $A>12$, 
the following parameterization \cite{BFG01} holds to a few percent for 
the whole range of very low energies up to $p_L \sim 100$ MeV/c: 

\begin{equation} \label{eq:rAdep}
R = \frac{7}{6}(1.840 + 1.120A^{1/3}) \; {\rm fm} \;\;\; .
\end{equation}
Equations (\ref{eq:sigma},\ref{eq:rAdep}) reproduce explicitly, 
for antiprotons, the dependence of the calculated total reaction 
cross sections on the three relevant parameters ($Z, A$ and energy).

\section{The antiproton - helium system}
\label{sec:He}

In this section I specialize to very light nuclear systems. 
Excluding the deuteron for which the optical model approach 
obviously is not well suited, the next targets to consider are the 
He isotopes \cite{BFG01}. The $(A-1)/A$ factor in 
Eq. (\ref{equ:potl}) is now included. Since the $2p$ and the 
$3d$ width data are incompatible with each other within an optical 
potential approach, as noted already in Ref. \cite{Sch91}, 
the $3d$ width data were excluded from the fitting procedure,
partly on the grounds that the $d$-wave contribution to the very 
low energy $\bar p$ annihilation cross sections is almost negligible
compared to the dominant $s$ and $p$ waves (see below).
The resulting density-folded optical potential, referred to as ($c$),
has the following parameters: 1.8 fm for the range parameter
of the two-body Gaussian interaction folded in with the He densities,
and a strength parameter $b_0 = -$ 0.26 + i2.07 fm. 
Again, its real part plays only a minor role.
The $2p$ level shifts and widths, as well as the $\bar p$ total annihilation
cross sections calculated using the optical potential ($c$) are shown in
Table \ref{tab:He1}, where the measured values are also given,
including a recent report of $\bar p$ annihilation on $^3$He \cite{Bia00c}.

\begin{table}
\caption{Measured $\bar p$ shifts ($\varepsilon$) and widths 
($\Gamma$) in eV \protect\cite{Sch91}, and $\bar p$ annihilation cross 
sections (in mb) on $^3$He \protect\cite{Bia00c}, $^4$He \protect
\cite{Zen99b} ($p_L$ in MeV/c). The calculations \protect\cite{BFG01} 
use potential (c)}
\label{tab:He1}
\begin{tabular}{lcccccc}
\hline
& & $\varepsilon _{2p}$ & $\Gamma_{2p}$ &
 $\sigma_{ann}$ (47.0)  &
$\sigma_{ann}$ (55.0)& $\sigma_{ann}$ (70.4)  \\
\hline
$^{3}$He & calc.& $-12$ & 33 & --- & 1038 &--- \\

$^{3}$He & exp.&$-17\pm 4$&$25\pm9 $& --- & 1850$\pm 700$& ---  \\
	
$^4$He&calc. & $-19$ &42& 1116 & --- &810  \\

$^4$He&exp. & $-18 \pm 2$ & $45 \pm 5$& $979 \pm
145$ & ---& $827 \pm 38$  \\
\hline
\end{tabular}
\end{table}

Protasov et al. \cite{PBL00} have recently fitted the two low-energy
$\bar p$ - $^4$He total annihilation cross sections listed in the
table, using the scattering-length approximation expressions for $s$, 
$p$ and $d$ waves \cite{CPZ97}. The input $p$- and $d$-wave 
Coulomb-modified scattering `lengths' were derived using the Trueman 
formula \cite{Tru61} 
from the $2p$ shift and $2p$ and $3d$ widths. The $s$-wave Coulomb-modified
scattering length $\tilde a_0$ was left as a fitting
parameter, since the $\bar p$ atomic $1s$ level
shift and width in He are not known experimentally.
Assuming a value of $1.0 \pm 0.5$ fm for Re~$\tilde a_0$, 
the annihilation cross sections were fitted by 
\begin{equation}
\label{eq:scPBL}
{\rm Im}~\tilde a_0 = - 0.36 \pm 0.03 ({\rm stat.})
^{+0.19}_{-0.11}
 ({\rm syst.}) \; {\rm fm} \;\;\; .
\end{equation} 
In contrast, the $\bar p$ - $^4$He potential ($c$), 
which is also fitted to essentially 
the same data set, yields numerically the following values:
\begin{equation}
\label{eq:scG'}
{\rm Re}~\tilde a_0 = 1.851 \; {\rm fm} \; , \;\;\;
{\rm Im}~\tilde a_0 = - 0.630 \; {\rm fm} \;\;\; .
\end{equation}
It is clear that the model independence claimed by 
Protasov et al. \cite{PBL00} is violated by this specific example 
and, therefore, the determination of $\tilde a_0$ is not model independent.
In order to study the origin of the above discrepancy, 
the partial wave contributions to the calculated cross section at 
$p_L = 57$ MeV/c are shown in Table \ref{tab:He2}. 
The most important contributions are due to the $s$- and 
$p$ waves, for which the two calculations disagree badly with each other. 
In particular, the $p$-wave contributions differ
from each other, although sharing practically the $\it same$
value of the Coulomb-modified $p$-wave `scattering length' (more
traditionally called `scattering volume') $\tilde a_1$. Whereas
at $-$20 keV, for the $2p$ atomic state, the $l=1$ $\bar p$ - $^4$He 
dynamics is well determined by $\tilde a_1$ alone, 
over the energy range of $1 - 3$ 
MeV, corresponding to the annihilation measurements, it depends on
more than just this `scattering length'. The effective
range term, and perhaps higher order terms in the effective range
expansion at low energies, become equally important. Indeed, it was 
verified for potential ($c$) that the variation of the Coulomb-modified
$l=1$ scattering phase shift is not reproduced in this energy range
by specifying the `scattering length' alone. Once the $p$-wave
contributions to the total annihilation cross section differ by as
much as is observed in the table, the $s$-wave 
contributions must also differ from each other. 
Therefore, the prediction for the Coulomb-modified $s$-wave scattering 
length using potential ($c$) is necessarily different from 
that of Ref. \cite{PBL00} which is based on an unjustified $p$-wave 
contribution. Such a difference would not occur for the 
$\bar p p$ system, which at the appropriate low energies is largely 
controlled by $s$ waves \cite{CPZ97}.

\begin{table}
\caption{Composition of $\bar p$ - $^{4}$He 
total annihilation cross section in mb}
\label{tab:He2}
\begin{tabular}{cccccc}
\hline
$p_L=57$ MeV/c  & $l = 0$ & $l = 1$ & $l = 2$&$l = 3$ & sum  \\
\hline
Protasov et al. \cite{PBL00} & 280.3 & 652.5 & 16.2 &  & 949  \\
potential ($c$) \cite{BFG01} & 395.8 & 500.3 & 49.8 & 1.5 & 949  \\
experiment \cite{Bia00a} &  &  &  &  & $915 \pm 32$  \\
\hline
\end{tabular}
\end{table}

Regarding the behavior of the $\bar p$-nucleus $s$-wave scattering 
lengths as function of $A$, say for $A>10$, 
the (plain strong-interaction) scattering lengths follow a simple 
geometrical picture, as borne out by 
a fit to $\bar p$-atom data \cite{Bat83} which has been recently 
updated \cite{BFG01}: 

\begin{equation}
\label{eq:sBat}
{\rm Re}~a_0 = (1.54 \pm 0.03)A^{0.311 \pm 0.005} \; {\rm fm} \; , \;\;\;
{\rm Im}~a_0 = -1.00 \pm 0.04 \; {\rm fm} \;\;\; . 
\end{equation}
Thus, $\rm Re ~a_0$ varies roughly as the nuclear radius $R$, whereas 
$\rm Im ~a_0$ is roughly constant over the periodic table. 
However, the Coulomb-modified scattering
lengths do not show such a clear geometrical picture \cite{BFG01}.

\section{Summary}
\label{sec:sum}

In conclusion, it was shown that the recently reported annihilation 
cross sections for $\bar p$ on $^4$He and Ne at 57 MeV/c are reproduced 
very well by strongly absorptive optical potentials which fit $\bar p$ 
atomic data. For these potentials, the asymptotic $ZA^{1/3}$ dependence 
of the total reaction cross section follows from the Coulomb focussing 
effect at very low energies. The full dependence on $E$, $Z$ and $A$ 
was derived using a Coulomb-modified strong absorption model.

\section*{ACKNOWLEDGEMENTS}
\label{sec:acknow}

I would like to acknowledge a longstanding collaboration on the topics here
reviewed with Drs. C.J. Batty and E. Friedman, and useful discussions with 
Drs. T. Bressani, A. Feliciello, F. Iazzi and T. Walcher. 
This research is partly supported by the DFG trilateral grant 243/51-2.


\begin{thebibliography}{bit99}


\bibitem{Zen99a}  A. Zenoni et al., Phys. Lett. B 461 (1999) 405.

\bibitem{Zen99b}  A. Zenoni et al., Phys. Lett. B 461 (1999) 413.

\bibitem{Bia00a}  A. Bianconi et al., Phys. Lett. B 481 (2000) 194.

\bibitem{Bia00b}  A. Bianconi, G. Bonomi, M.P. Bussa, E. Lodi Rizzini, 
L. Venturelli and A. Zenoni, Phys. Lett. B 483 (2000) 353.
	
\bibitem{GFB00}   A. Gal, E. Friedman and C.J. Batty, Phys. Lett. B 
491 (2000) 219. 

\bibitem{BFG01}  C.J. Batty, E. Friedman and A. Gal, Nucl. Phys. A 
689 (2001) 721.

\bibitem{BFG97}  C.J. Batty, E. Friedman and A. Gal, Phys. Rep. 287 
(1997) 385.

\bibitem{BFG95}  C.J. Batty, E. Friedman and A. Gal, Nucl. Phys. A 
592 (1995) 487.

\bibitem{SMC79}  K. Stricker, H. McManus and J.A. Carr, 
Phys. Rev. C 19 (1979) 929.

\bibitem{FGJ91}  E. Friedman et al., Phys. Lett. B 257 (1991) 17.

\bibitem{FGa99a}  E. Friedman and A. Gal, Phys. Lett. B 459 (1999) 43.
	
\bibitem{FGa99b}  E. Friedman and A. Gal, Nucl. Phys. A 658 (1999) 345.

\bibitem{Trz01}  A. Trzcinska et al. (for experiment PS209), 
Proc. LEAP 2000, Nucl. Phys. A (in press). 

\bibitem{Che75}  S.C. Cheng et al., Nucl. Phys. A 254 (1975) 381. 

\bibitem{Sch91}  M. Schneider et al., Z. Phys. A 338 (1991) 217.

\bibitem{Bot01}  E. Botta (for the OBELIX experiment), Proc. LEAP 2000, 
Nucl. Phys. A (in press). 

\bibitem{Bal86}  F. Balestra et al., Nucl. Phys. A 452 (1986) 573. 

\bibitem{Bla54}  J.S. Blair, Phys. Rev. 95 (1954) 1218.

\bibitem{Schiff}  L.I. Schiff, Quantum Mechanics, third edition, 
McGraw Hill, New York, 1968. 

\bibitem{Bia00c}A. Bianconi et al., Phys. Lett. B 492 (2000) 254.

\bibitem{PBL00}  K.V. Protasov, G. Bonomi, E. Lodi Rizzini and A. Zenoni, 
Eur. Phys. J. A 7 (2000) 429.

\bibitem{CPZ97}  J. Carbonell, K.V. Protasov and A. Zenoni, Phys. Lett. B 
397 (1997) 345.

\bibitem{Tru61}  T.L. Trueman, Nucl. Phys. 26 (1961) 57.

\bibitem{Bat83}  C.J. Batty, Nucl. Phys. A 411 (1983) 399. 

\end{thebibliography}
\end{document}